\documentclass[conference]{IEEEtran}
\IEEEoverridecommandlockouts
\usepackage{cite}
\usepackage{amsmath,amssymb,amsfonts}
\usepackage{algorithm,subcaption}
\usepackage{algorithmic}
\usepackage{graphicx}
\usepackage{textcomp}
\usepackage{xcolor}
\def\BibTeX{{\rm B\kern-.05em{\sc i\kern-.025em b}\kern-.08em    
T\kern-.1667em\lower.7ex\hbox{E}\kern-.125emX}}
\usepackage{hyperref}
\begin{document}

\title{Period-conscious Time-series Reconstruction under Local Differential Privacy}

\author{
  Yaxuan Wang$^{1*}$, Tianxin Li$^{2*}$, Enji Liang$^1\dagger$, Yue Fu$^{1}$, and Yanran Wang$^{1\dagger}$
  \thanks{
    This paper has been accepted by the ICME 2026 Workshop. * These authors contributed equally to this work. $^{\dagger}$ Corresponding authors.}
  \\
  \IEEEauthorblockA{
    1. Taiyuan University of Technology, China\\
    2. University of Michigan, Ann Arbor, USA\\
    Emails: 2023005749@link.tyut.edu.cn, ltianxin@umich.edu, liangenji@tyut.edu.cn, fuyue@tyut.edu.cn, 2023006418@link.tyut.edu.cn
  }
}
\maketitle

\begin{abstract}
Periodic patterns are fundamental cues in multimedia signals and systems, including repetitive motion in video (e.g., gait cycles), rhythmic and pitch-related structure in audio, and recurring textures in image sequences. When such user-generated streams are collected from edge devices, local differential privacy (LDP) is appealing because it perturbs data before upload; however, the injected noise can corrupt spectral peaks and induce phase drift, making period estimation unreliable and degrading reconstruction quality. 
We propose \textbf{CPR} (\textit{Cycle and Phase Recovery}), a period-aware reconstruction framework for periodic time series under LDP. CPR performs multi-scale period probing and multi-consensus selection to suppress noise-induced spectral interference, then aggregates perturbed samples at matched within-cycle phase positions to stabilize phase alignment across cycles. To recover the underlying per-phase values, CPR combines EM-based denoising with kernel density estimation, improving robustness under tight privacy budgets.
Experiments on two real-world periodic datasets demonstrate that CPR better preserves periodic structure and consistently achieves lower reconstruction error than representative LDP baselines, especially in the low-$\epsilon$ regime.
\end{abstract}

%\begin{IEEEkeywords}
%Local differential privacy, periodic time series, phase-aware aggregation, time-series reconstruction, privacy-preserving multimedia analytics
%\end{IEEEkeywords}

\section{Introduction}
\label{sec:intro}

Periodic time series are ubiquitous in multimedia and multimodal sensing pipelines. Many multimedia tasks rely on stable periodic structure: repetitive motion patterns in video for action and gait analysis \cite{ref24}\cite{ref28}, pitch- and rhythm-related trajectories in audio for synthesis and enhancement \cite{ref25}\cite{ref29}, and recurring visual patterns across frames for texture analysis and reconstruction \cite{ref26}. In practice, these periodic cues often appear not only in raw signals, but also in compact time-varying descriptors derived from multimedia content---e.g., pose/keypoint trajectories, motion energy, optical-flow statistics, pitch contours, or onset-strength envelopes---which are widely used for efficient indexing, summarization, synchronization, and downstream recognition. Beyond classic multimedia, similar periodic streams also arise in human-centric sensing (e.g., physiological monitoring) and other continuous modalities, where preserving cycle-level structure is critical for reliable modeling and decision making \cite{ref1,ref2}.

At the same time, collecting and sharing such streams raises serious privacy risks. Continuous video-, audio-, and sensor-derived time series can reveal identities, routines, and sensitive attributes (e.g., health conditions or behavioral patterns), even when the shared data are lower-dimensional features rather than raw media. Local Differential Privacy (LDP) provides a strong protection model by perturbing each user’s data \emph{before} transmission, making privacy independent of the trustworthiness of the collector. Formally, for any two inputs \(t\) and \(t'\) and any output \(t^*\),
\[
\Pr[M(t) = t^*] \le e^{\epsilon} \cdot \Pr[M(t') = t^*],
\]
\cite{ref16,ref3,ref11,ref19} where a smaller \(\epsilon\) enforces stronger privacy. This local model is particularly appealing for edge and mobile deployments with continuous data upload, where raw streams are hard to protect and auditing centralized processing is challenging.

However, LDP noise can be especially destructive for periodic signals, particularly under strict budgets (\(\epsilon < 1\)) \cite{ref3,ref5}. Because perturbation is injected at the sample level, it can (i) pollute the spectrum and blur dominant frequencies, making peak-based period estimation unstable; (ii) amplify apparent misalignment across cycles (often described as ``phase drift''), where small errors in period detection or boundary localization accumulate into noticeable within-cycle offset; and (iii) exacerbate boundary artifacts in streaming publication, where noise accumulation and windowed budget splitting can break cross-cycle symmetry \cite{ref12}. In multimedia settings, such failures translate into unstable cycle estimates, degraded reconstruction, and reduced utility for applications such as motion analysis, rhythmic modeling, or periodic texture reconstruction. Visually or perceptually, boundary effects can manifest as discontinuities between repeated segments---e.g., audible jitter in periodic audio or visible seams in reconstructed frame sequences---highlighting the need for reconstruction procedures that explicitly respect cycle/phase structure rather than only minimizing pointwise error.

Existing LDP solutions do not adequately address this issue. Budget allocation strategies for streams (e.g., LBD \cite{ref7}) can improve efficiency, but they are not designed to robustly recover heterogeneous periodic structures common in real-world multimedia, such as variable period lengths, tempo drift, and multi-period superposition \cite{ref13,ref14}. Other reconstruction approaches (e.g., compressed sensing based methods) may rely on assumptions that are difficult to guarantee in practical deployments \cite{ref15}, limiting robustness for high-quality reconstruction. These gaps motivate a period-conscious view of LDP reconstruction: the goal is not only to denoise, but to \emph{stabilize the dominant cycle} and \emph{align phases} so that the reconstructed stream preserves the structure most relevant to periodic multimedia analysis.

To address these challenges, we propose \textbf{CPR} (\textit{Cycle and Phase Recovery}), a period-conscious reconstruction framework for periodic time series under LDP. Real-world signals can be complex with many rhythms. However, in strong LDP noise, small rhythms are usually completely covered by noise spikes. CPR focuses on finding the most stable period because it is the most useful part of the signal that can still be saved. This approach is practical and ensures the best possible recovery under high privacy protection. The main contribution of this paper is three-fold:
\begin{itemize}
    \item \textbf{A period-conscious reconstruction problem formulation for LDP-protected multimedia streams.} We highlight how LDP noise corrupts periodic structure through spectral pollution, phase drift, and boundary effects in continuous publishing, motivating reconstruction methods that explicitly preserve cycle/phase information.
    \item \textbf{Multi-scale, multi-consensus period identification under LDP.} CPR probes candidate periods across multiple window sizes and selects reliable cycles via cross-window agreement, suppressing spurious peaks caused by privacy noise and supporting heterogeneous periodic patterns.
    \item \textbf{Phase-aware aggregation and density-based reconstruction.} Given the estimated period, CPR groups perturbed observations by within-cycle phase, stabilizes cross-cycle alignment, and reconstructs per-phase values using an EM-then-KDE strategy, yielding improved accuracy and robustness under tight privacy budgets.

\end{itemize}

\section{Preliminaries}
\label{sec:prelim}

\subsection{Notation and normalization}
\label{sec:prelim-norm}
We consider a length-$n$ time series
$X^{\mathrm{raw}}=\{x^{\mathrm{raw}}_t\}_{t=1}^{n}$.
Each device maps values to a bounded domain for privacy mechanisms defined on $[0,1]$:
\begin{equation}
\label{eq:norm}
x_t=\mathrm{Normalize}(x^{\mathrm{raw}}_t,[0,1])\in[0,1],\qquad
X=\{x_t\}_{t=1}^{n}.
\end{equation}

\subsection{Approximate periodicity and dominant period}
\label{sec:prelim-period}
A standard proxy for ``how periodic'' a series is at lag $T$ is the mean squared shift discrepancy
\begin{equation}
\label{eq:period-loss}
\Delta_T(X)=\frac{1}{n-T}\sum_{t=1}^{n-T}(x_t-x_{t+T})^2.
\end{equation}
A dominant period is a $T$ (within an admissible range) for which $\Delta_T(X)$ is comparatively small.

\subsection{$w$-event local differential privacy and budget splitting}
\label{sec:prelim-wldp}
We adopt $w$-event LDP for streams \cite{ref7}. Let $I=\{t,\dots,t+w-1\}$ be any contiguous window of length $w$, and let $M_I$ be the joint release over $I$.
A mechanism satisfies $\epsilon$-$w$-event LDP if for any two window inputs $u,v\in[0,1]^w$ and any measurable output set $\mathcal{O}$,
\begin{equation}
\label{eq:w-event-ldp}
\Pr[M_I(u)\in\mathcal{O}] \le e^\epsilon \Pr[M_I(v)\in\mathcal{O}].
\end{equation}
If each per-time mechanism is $\epsilon_0$-LDP, then by sequential composition any length-$w$ window is $(w\epsilon_0)$-LDP. Hence we set
\begin{equation}
\label{eq:eps-split}
\epsilon_0=\epsilon/w,
\end{equation}
so the stream satisfies \eqref{eq:w-event-ldp}.

\subsection{Square Wave (SW) local randomizer}
\label{sec:prelim-sw}
We use the SW mechanism \cite{ref20} as the per-time local randomizer. Given $x\in[0,1]$, SW outputs $y\in[0,1]$ with a two-level conditional density:
\begin{equation}
\label{eq:sw-density}
f_{\mathrm{SW}}(y\mid x)=
\begin{cases}
p, & y\in I_x,\\
q, & y\in[0,1]\setminus I_x,
\end{cases}
\end{equation}
where $I_x$ is a length-$2b$ interval ``centered'' at $x$ but shifted to stay inside $[0,1]$:
\[
I_x=
\begin{cases}
[0,2b], & x<b,\\
[x-b,x+b], & b\le x\le 1-b,\\
[1-2b,1], & x>1-b.
\end{cases}
\]
Parameters $(b,p,q)$ are determined by $\epsilon_0$ (with $p/q=e^{\epsilon_0}$) as in \cite{ref20}:
\[
b=\frac{\epsilon_0 e^{\epsilon_0}-e^{\epsilon_0}+1}{2e^{\epsilon_0}\bigl(e^{\epsilon_0}-1-\epsilon_0\bigr)},\qquad
p=\frac{e^{\epsilon_0}}{2b e^{\epsilon_0}+1},\qquad
q=\frac{1}{2b e^{\epsilon_0}+1}.
\]
Applying SW independently at each time $t$ yields the privatized stream $X'=\{x'_t\}_{t=1}^n$.

\subsection{FFT peak index to period conversion}
\label{sec:prelim-fft}
For a length-$N$ FFT, a positive-frequency index $k\in\{1,\dots,\lfloor N/2\rfloor\}$ corresponds to an approximate period (in samples)
\begin{equation}
\label{eq:fft2period}
T(k)=\mathrm{round}\!\left(\frac{N}{k}\right).
\end{equation}

\subsection{SW decoding via EM on a discretized grid}
\label{sec:prelim-em}
To denoise SW outputs, we discretize $[0,1]$ into $B$ grid points
$v_b=(b-\tfrac{1}{2})/B$ and estimate a latent pmf $\pi_b$ via EM.
Given observations $\{y_j\}_{j=1}^m$, the E/M updates are
\begin{align}
\label{eq:em-e}
r_{j,b} &=
\frac{\pi_{b}^{(t)} f_{\mathrm{SW}}(y_j\mid v_b)}
{\sum_{b'=1}^B \pi_{b'}^{(t)} f_{\mathrm{SW}}(y_j\mid v_{b'})},\\
\label{eq:em-m}
\pi_{b}^{(t+1)} &= \frac{1}{m}\sum_{j=1}^{m} r_{j,b}.
\end{align}
We convert responsibilities to denoised pseudo-samples by posterior means:
\begin{equation}
\label{eq:em-pseudo}
\hat{y}_j=\sum_{b=1}^{B} r_{j,b}v_b.
\end{equation}

\subsection{KDE and mode extraction}
\label{sec:prelim-kde}
Given samples $\{z_j\}_{j=1}^m\subset[0,1]$, a Gaussian-kernel KDE is
\begin{equation}
\label{eq:kde}
\hat{f}(x)=\frac{1}{mh}\sum_{j=1}^m
\exp\!\left(-\frac{(x-z_j)^2}{2h^2}\right),
\end{equation}
with bandwidth $h$ (e.g., Silverman's rule). We use the KDE mode
$x^\star=\arg\max_{x\in[0,1]}\hat{f}(x)$ as a robust point estimate.

\section{Periodic Time-Series Aware Reconstruction}
\label{sec:method}

\subsection{Problem Statement}
\label{sec:setting}
Each device holds a normalized stream $X=\{x_t\}_{t=1}^{n}\subset[0,1]$ and releases only its privatized counterpart
$X'=\{x'_t\}_{t=1}^{n}$ by applying the SW local randomizer independently at each time with $\epsilon_0=\epsilon/w$
(Sec.~\ref{sec:prelim-wldp}--\ref{sec:prelim-sw}).
The server observes $X'$ and aims to recover the \emph{dominant periodic component} and reconstruct a denoised stream
$\hat{X}$ that preserves the major cyclic pattern while suppressing LDP-induced noise.

A convenient interpretation is that the original stream admits a dominant (possibly approximate) cycle of length $T$:
\begin{equation}
\label{eq:model}
x_t \approx r_{((t-1)\bmod T)+1} + \xi_t,
\end{equation}
where $r\in[0,1]^T$ is an unknown template and $\xi_t$ captures non-periodic deviations. The main difficulties are:
(i) SW noise significantly flattens spectra and introduces spurious peaks; (ii) real streams are only approximately periodic, so
single-scale FFT peak picking is unstable; and (iii) even with the correct $T$, phase-aligned samples remain noisy and require
privacy-aware denoising.

\subsection{System framework}
\begin{figure}
    \centering
    \includegraphics[width=\linewidth]{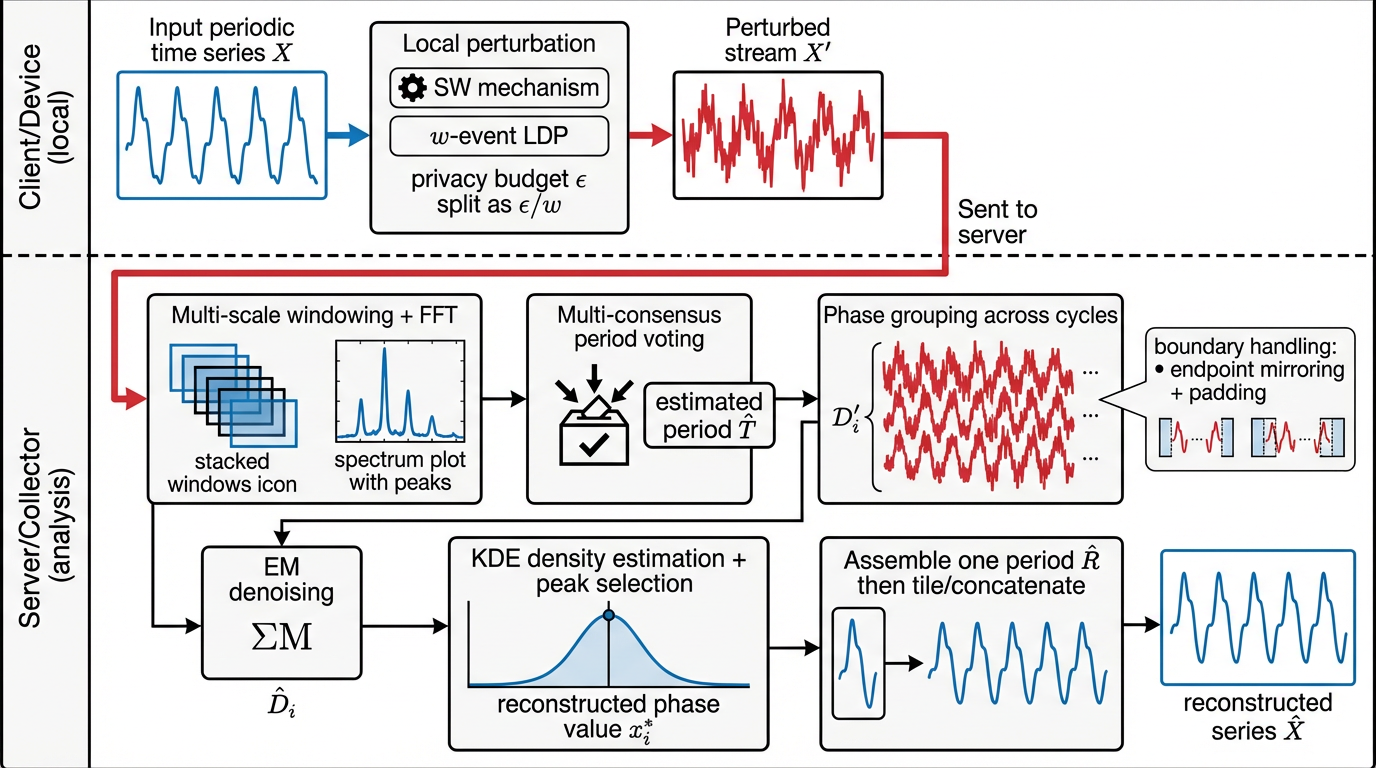}
    \caption{System overview. Devices locally privatize the stream and upload only $X'$. The server runs CPR to estimate the dominant period and reconstruct a representative cycle.}
    \label{fig:framework}
\end{figure}

Figure~\ref{fig:framework} summarizes the pipeline. The device-side computation is a single pass SW perturbation (line~1 in Alg.~\ref{alg:CPR}),
which guarantees $\epsilon$-$w$-event LDP by composition (Sec.~\ref{sec:prelim-wldp}).
All subsequent steps are server-side post-processing on $X'$ and do not incur additional privacy loss.

At a high level, CPR separates the task into:
\emph{cycle recovery} (estimate $\hat{T}$ robustly under LDP noise) and
\emph{phase recovery} (estimate the cycle template $\hat{R}$ by aggregating phase-aligned noisy samples and denoising them).

\subsection{Cycle recovery: robust dominant period estimation}
\label{sec:period-detect}
Real-world signals can be complex with many rhythms. However, in strong LDP noise, small rhythms are usually completely covered by noise spikes. CPR focuses on finding the most stable period because it is the most useful part of the signal that can still be saved. This approach is practical and ensures the best possible recovery under high privacy protection. 
\subsubsection{Multi-scale spectral candidate generation}
\label{sec:multiscale}
Rather than relying on a single FFT over the full stream, CPR probes periodicity across multiple window sizes $S$.
This mitigates spectral leakage and local nonstationarity: short windows adapt to drift, while long windows improve frequency resolution.

For each $s\in S$, we slide length-$s$ windows with hop $h_s=\lfloor s/2\rfloor$.
Given a window $z\in\mathbb{R}^s$, we apply standard FFT-friendly preprocessing to stabilize peak detection:
\begin{equation}
\label{eq:fft-preprocess}
z \leftarrow z - \mathrm{mean}(z)\mathbf{1}\qquad
(\text{optional: } z \leftarrow z \odot w_{\mathrm{Hann}}),
\end{equation}
then compute the power spectrum on a zero-padded length $N_s=2^{\lceil\log_2 s\rceil}$ FFT.
We extract the top-$L$ non-DC peaks $k$ and map each peak to a period candidate $T(k)$ via Eq.~\eqref{eq:fft2period},
discarding candidates outside $[T_{\min},T_{\max}]$.

\subsubsection{Time-domain validation to suppress spurious FFT peaks}
\label{sec:rep-validate}
FFT peaks under strong LDP noise can be unreliable (e.g., random local maxima or harmonics).
We therefore validate each candidate period $T$ using \emph{repeatability} in the time domain.
Within the current window, we form up to $M=\min\{3,\lfloor s/T\rfloor\}$ consecutive length-$T$ segments and compute the mean adjacent cosine similarity of de-meaned segments. If $\lfloor s/T\rfloor<2$ (insufficient repeats), we treat the candidate as invalid in that window.

This step plays two roles: (i) it rejects peaks that do not correspond to a truly repeating shape, and (ii) it helps disambiguate harmonics, since a false harmonic often yields weaker segment-to-segment consistency after de-meaning. For each window we keep the candidate $T^\star$ with the highest repeatability score, producing a set of window-level estimates
at each scale $s$.

\subsubsection{Robust aggregation within scale and cross-scale consensus}
\label{sec:consensus}
At a fixed scale $s$, CPR aggregates window-level decisions using medians:
$T_s=\mathrm{median}(\{T^\star\})$ and $q_s=\mathrm{median}(\{\mathrm{Rep}\})$.
The median is resilient to outlier windows (e.g., abrupt changes or noisy peak selections).

Across scales, CPR performs tolerance-based voting.
For each integer $T\in[T_{\min},T_{\max}]$, define $\delta_T=\max(1,\lceil\tau T\rceil)$ and count supporting scales
$\#\{s\in S:\ |T_s-T|\le \delta_T\}$.
We additionally enforce agreement across both short and long scales (split at the median of $S$) to avoid selecting a period
that appears only at one resolution.
Among candidates with maximal support, we break ties by higher average repeatability, and if still tied we select the smaller period
(to favor the fundamental over harmonics when evidence is comparable).
The resulting $\hat{T}$ is the detected dominant period.

\subsection{Phase recovery: privacy-aware template reconstruction}
\label{sec:phase-recon}

\subsubsection{Phase grouping and boundary stabilization}
\label{sec:phase-group}
Given $\hat{T}$, CPR pools samples that correspond to the same phase in the cycle.
To reduce boundary effects (unequal counts near the start/end), we mirror-pad $X'$ by $p=\hat{T}-1$ samples on both sides
and form the padded stream $\tilde{X}'$.
Let $M=\lfloor |\tilde{X}'|/\hat{T}\rfloor$.
For each phase $i\in\{1,\dots,\hat{T}\}$ we collect
$D'_i=\{\tilde{x}'_{i+m\hat{T}}\}_{m=0}^{M-1}$.

Intuitively, if the true signal is dominantly periodic as in \eqref{eq:model}, then each $D'_i$ is a batch of noisy privatized
observations of (approximately) the same latent value $r_i$, so pooling across cycles increases the effective sample size.

\subsubsection{SW-aware denoising and robust point estimation}
\label{sec:phase-denoise}
Because SW produces a characteristic two-level conditional density (Sec.~\ref{sec:prelim-sw}), naive averaging of $D'_i$
does not optimally invert the randomization.
We therefore apply the SW-tailored EM decoding on a discretized grid (Sec.~\ref{sec:prelim-em}) to obtain denoised pseudo-samples
$\hat{D}_i$ (Eq.~\eqref{eq:em-pseudo}).
We then summarize $\hat{D}_i$ by the mode of a 1D KDE (Sec.~\ref{sec:prelim-kde}), yielding the reconstructed phase value $x_i^\star$.
Using a mode (rather than a mean) improves robustness when the decoded distribution is skewed or mildly multi-modal, which can occur
under approximate periodicity or when sporadic events perturb a subset of cycles.

The reconstructed template cycle is $\hat{R}=(x_1^\star,\dots,x_{\hat{T}}^\star)$, and the final reconstructed stream
$\hat{X}$ is obtained by tiling $\hat{R}$ and cropping to length $n$.

\subsection{Privacy and post-processing}
\label{sec:privacy-post}
CPR performs only post-processing on $X'$. Since $X'$ is generated locally under $\epsilon_0=\epsilon/w$ per-time privacy,
the overall release satisfies $\epsilon$-$w$-event LDP by composition (Sec.~\ref{sec:prelim-wldp}), and the cycle/phase recovery
steps do not consume additional privacy budget. This separation is crucial in practice: devices run a fixed lightweight perturbation,
while the server can iterate on reconstruction logic without affecting privacy guarantees.

\begin{algorithm}[t]
\caption{\textbf{CPR: Cycle and Phase Recovery}}
\label{alg:CPR}
\begin{algorithmic}[1]
\REQUIRE raw stream $X^{\mathrm{raw}}$ (length $n$), total budget $\epsilon$, event window $w$, probing scales $S$, admissible $[T_{\min},T_{\max}]$, peak count $L$, tolerance $\tau$, EM grid size $B$
\ENSURE reconstructed stream $\hat{X}$
\STATE $X\!\leftarrow\!\mathrm{Normalize}(X^{\mathrm{raw}},[0,1])$; $\epsilon_0\!\leftarrow\!\epsilon/w$; $x'_t\!\leftarrow\!\mathrm{SW}(x_t;\epsilon_0)$ for all $t$ 
\STATE $\mathcal{C}\leftarrow\emptyset$
\FORALL{$s\in S$}
  \STATE Scan length-$s$ windows of $X'$; in each window, select $T^\star$ from top-$L$ FFT peaks by maximizing $\mathrm{Rep}(\cdot,T)$; keep only $T^\star\!\in[T_{\min},T_{\max}]$
  \STATE $T_s\leftarrow\mathrm{median}(\{T^\star\})$; $q_s\leftarrow\mathrm{median}(\{\mathrm{Rep}\})$; $\mathcal{C}\leftarrow \mathcal{C}\cup\{(T_s,q_s,s)\}$
\ENDFOR
\STATE $\hat{T}\leftarrow \mathrm{ConsensusVote}(\mathcal{C},\tau,S)$
\STATE $\tilde{X}'\leftarrow \mathrm{MirrorPad}(X',\hat{T}-1)$; $M\leftarrow \left\lfloor |\tilde{X}'|/\hat{T}\right\rfloor$
\FOR{$i=1$ to $\hat{T}$}
  \STATE $D'_i\leftarrow\{\tilde{x}'_{i+m\hat{T}}\}_{m=0}^{M-1}$; $\hat{D}_i\leftarrow \mathrm{EM\_SW}(D'_i,B,\epsilon_0)$
  \STATE $x_i^\star \leftarrow \mathrm{ModeKDE}(\hat{D}_i)$
\ENDFOR
\STATE $\hat{R}\leftarrow(x_1^\star,\dots,x_{\hat{T}}^\star)$; $\hat{X}\leftarrow \mathrm{TileCrop}(\hat{R},n)$; \RETURN $\hat{X}$
\end{algorithmic}
\end{algorithm}

\section{Experimental Setup and Results}
\label{3}
\subsection{Hardware and Software Environment}
\textbf{Setup}: All experiments were conducted on a workstation equipped with an Intel Core i7-13650HX processor and 16 GB of RAM, running Windows 11.

\textbf{Dataset}: We use the DARWIN Daily Meridian Longitude dataset referred to as \textbf{Darwin}\footnote{https://archive.ics.uci.edu/dataset/732/darwin}, which contains handwritten samples collected from participants. Additionally, we include the Turkish Music Emotion dataset referred to as \textbf{Music}\footnote{https://archive.ics.uci.edu/dataset/862/turkish+music+emotion}, the Raisin dataset referred to as \textbf{Raisin}\footnote{https://archive.ics.uci.edu/dataset/850/raisin}, and the Crowdsourced Image dataset referred to as \textbf{Crowdsourced}\footnote{https://archive.ics.uci.edu/dataset/400/crowdsourced+mapping}.To simulate periodic multimedia streams, we extracted continuous numerical features (e.g., pressure trajectories in Darwin, frequency envelopes in Music) and concatenated repeated segments with minor synthetic jitter to represent the quasi-periodicity typical of real-world sensing data.

\textbf{Data Preparation:}While the selected datasets cover diverse domains, we processed them into periodic streams to evaluate CPR's reconstruction performance. For each dataset, we selected the primary numerical feature as the target  periodic signal. 

\textbf{Comparison Methods}: We compare six representative methods and evaluate their reconstruction performance using cosine distance. The methods include: \textbf{Laplace}~\cite{ref22}, which applies the standard Laplace mechanism at each node with an appropriate noise scale followed by smoothing; \textbf{SW}, which directly perturbs the data using the SW algorithm and treats the perturbed time series as the reconstruction; two enhanced SW variants, \textbf{SW\_moving} and \textbf{SW\_filter}, which apply moving-average smoothing and filter-based smoothing, respectively, to further improve reconstruction quality; \textbf{LBD}~\cite{ref7}, which strictly follows the original implementation and dynamically allocates privacy budgets based on the magnitude of differences and potential publishing errors; and finally our method \textbf{CPR}, which detects cycles using multi-window combinations and aggregates phase points via Gaussian kernel density estimation to recover the periodic structure.

\subsection{Measurement of Periodic Accuracy}
\label{subsec:periodic-accuracy}

We evaluate the accuracy of periodic detection under the $w$-event LDP setting by varying the overall privacy budget and the window size. Specifically, we sweep the privacy budget
$\epsilon \in \{0.5, 1.0, 1.5, \ldots, 5.0\}$ and the window size $w \in \{5, 10, 15, 20, 25\}$.
Following the $w$-event LDP design, the privacy budget is evenly allocated within each window so that each of the $w$ event reports is perturbed using a per-event budget
\[
\text{eps} \;=\; \frac{\epsilon}{w},
\]
which ensures that the entire window satisfies the target $\epsilon$-LDP guarantee under sequential composition.
For every $(\epsilon, w)$ configuration, we run 100 independent trials and report the periodic detection accuracy as the percentage of trials in which the periodicity is correctly identified. For conciseness, Table~\ref{tab:periodic-acc} reports representative results for $\epsilon \in \{1,2,3,4,5\}$ (the omitted intermediate budgets follow the same overall trend).

\begin{table}[]
    \centering
    \caption{Periodic detection accuracy (\%) under different privacy budgets $\epsilon$ and window sizes $w$ for four datasets (100 trials per setting). The per-event budget is $\text{eps}=\epsilon/w$.}
    \label{tab:periodic-acc}
    \vspace{1mm}
    \small
    \setlength{\tabcolsep}{5pt}
    \begin{tabular}{l c c c c c c}
        \hline
        \textbf{Dataset} & $w$ & $\epsilon=1$ & $\epsilon=2$ & $\epsilon=3$ & $\epsilon=4$ & $\epsilon=5$ \\
        \hline
        \textbf{Darwin}        & 5  & 19 & 32 & 75 & 97 & 98 \\
                             & 10 & 16 & 25 & 28 & 50 & 77 \\
                             & 15 & 14 & 14 & 15 & 17 & 39 \\
                             & 20 & 12 & 13 & 15 &  8 & 22 \\
                             & 25 & 12 & 13 & 13 & 14 & 14 \\
        \hline
        \textbf{Music}         & 5  &  6 & 35 & 88 & 100 & 100 \\
                             & 10 &  2 &  3 &  9 &  35 &  65 \\
                             & 15 &  2 &  2 &  3 &   8 &  16 \\
                             & 20 &  1 &  2 &  3 &   4 &   7 \\
                             & 25 &  1 &  1 &  1 &   3 &   7 \\
        \hline
        \textbf{Raisin}        & 5  & 18 & 25 & 35 & 43 & 70 \\
                             & 10 & 21 & 21 & 28 & 25 & 36 \\
                             & 15 & 17 & 21 & 21 & 23 & 30 \\
                             & 20 & 13 & 16 & 20 & 21 & 26 \\
                             & 25 & 11 & 12 & 16 & 20 & 22 \\
        \hline
        \textbf{Crowdsourced}  & 5  &  8 & 21 & 29 & 49 & 70 \\
                             & 10 &  6 & 18 & 13 & 21 & 67 \\
                             & 15 &  5 & 10 & 10 & 17 & 21 \\
                             & 20 &  4 &  6 &  6 &  8 & 15 \\
                             & 25 &  0 &  1 &  4 &  6 &  8 \\
        \hline
    \end{tabular}
\end{table}

\textbf{Impact of the privacy budget $\epsilon$.}
For any fixed window size $w$, increasing $\epsilon$ generally improves periodic detection accuracy across all datasets. This behavior is expected under LDP: a larger $\epsilon$ reduces the amount of randomization injected into each report, thereby preserving periodic features more faithfully. The improvement is particularly pronounced when the periodic signal is strong and the window is small; for example, at $w=5$ the accuracy on \textbf{Darwin} increases from 19\% ($\epsilon=1$) to 98\% ($\epsilon=5$), and on \textbf{Music} from 6\% ($\epsilon=1$) to 100\% ($\epsilon\ge 4$). Minor non-monotonic fluctuations may appear at some settings (e.g., \textbf{Darwin} at $w=20$), which can occur due to finite-sample variability and the stochastic nature of local perturbation, but the dominant trend remains positive with increasing $\epsilon$.

\textbf{Impact of the window size $w$.}
For a fixed overall budget $\epsilon$, smaller window sizes typically yield better accuracy. The key reason is the per-event budget $\text{eps}=\epsilon/w$: increasing $w$ reduces $\text{eps}$ and therefore increases the perturbation magnitude per report. As a concrete illustration, when $\epsilon=5$, using $w=5$ allocates $\text{eps}=1.0$ per event, whereas $w=25$ allocates only $\text{eps}=0.2$; correspondingly, the accuracy drops from 98\% to 14\% on \textbf{Darwin} and from 100\% to 7\% on \textbf{Music}. This indicates that overly large windows can make periodic signatures difficult to recover under the same overall privacy constraint.

\textbf{Dataset-dependent sensitivity.}
The magnitude of the gains depends on the dataset: \textbf{Darwin} and \textbf{Music} benefit the most from larger $\epsilon$ and smaller $w$, suggesting clearer periodic structures that become detectable once LDP noise is sufficiently reduced. In contrast, \textbf{Raisin} and \textbf{Crowdsourced} show more gradual improvements, consistent with weaker or more heterogeneous periodic patterns where the same privacy-induced noise causes a larger relative degradation.

In summary, periodic detection under $w$-event LDP is strongly governed by the effective per-event budget $\epsilon/w$: accurate identification is achievable when $\epsilon$ is sufficiently large and $w$ is kept small, whereas large windows under tight privacy budgets can substantially reduce detection reliability.

\subsection{Comparison of Data Reconstruction Effects}

 Experiments are then conducted for different values of $\epsilon$= \{0.5, 1.0, 1.5, 2.0, 2.5, 3.0, 3.5, 4.0, 4.5, 5.0\} with $w$=5. For each method, we compare the cosine distance metric against the true values. All distribution data are normalized and interpolated to a unified coordinate system to ensure a fair comparison, measuring the overall similarity between the reconstructed sequence and the original sequence.
\begin{figure*}[]
    \centering
    % 第一张图: Darwin
    \begin{subfigure}[b]{0.24\textwidth}
        \centering
\includegraphics[width=\linewidth]{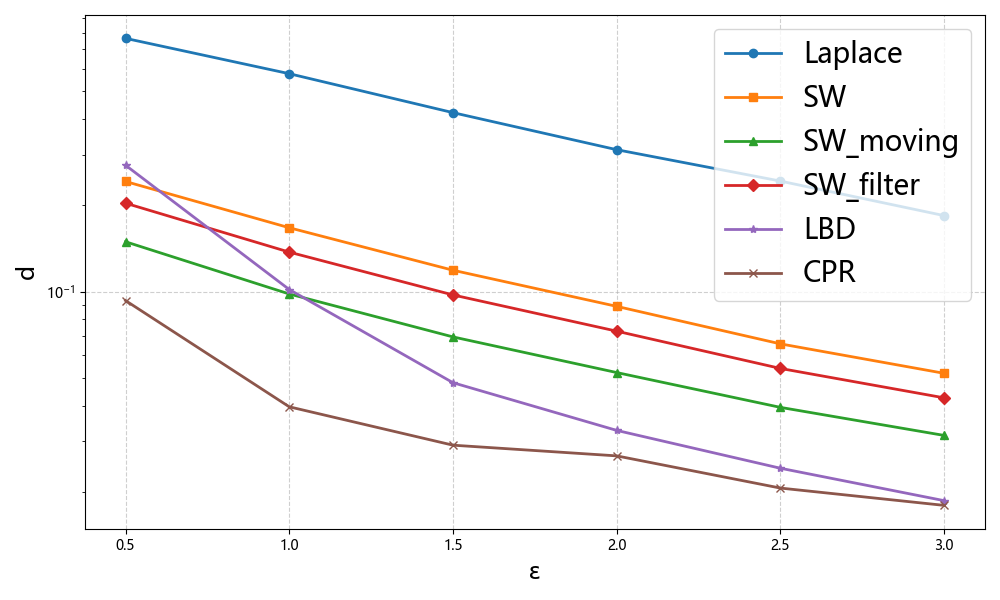}
        \caption{Darwin}
        \label{fig:darwin}
    \end{subfigure}
    \hfill
    % 第二张图: Music
    \begin{subfigure}[b]{0.24\textwidth}
        \centering
        \includegraphics[width=\linewidth]{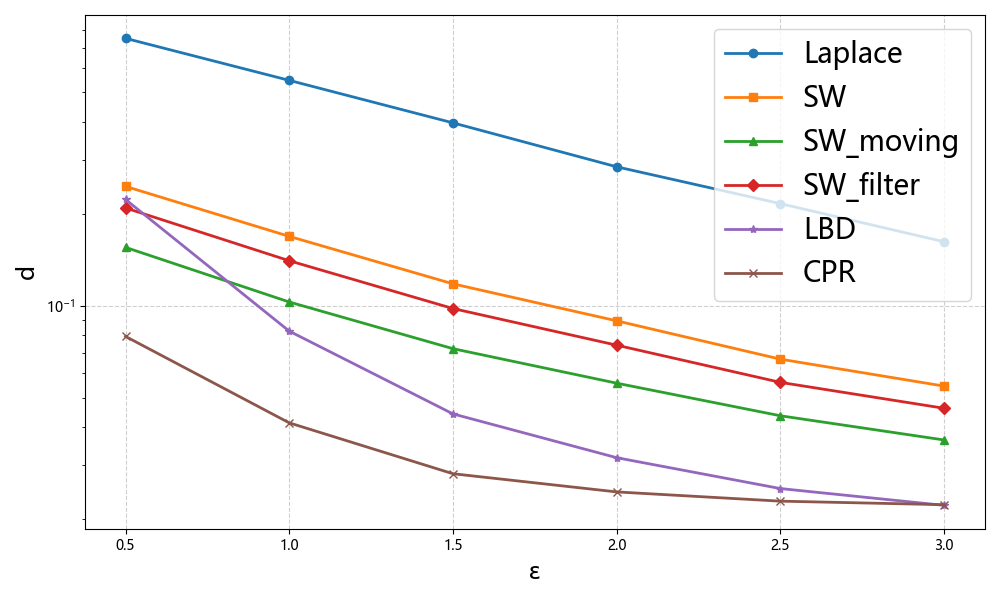}
        \caption{Music}
        \label{fig:music}
    \end{subfigure}
    \hfill
    % 第三张图: Raisin
    \begin{subfigure}[b]{0.24\textwidth}
        \centering
        \includegraphics[width=\linewidth]{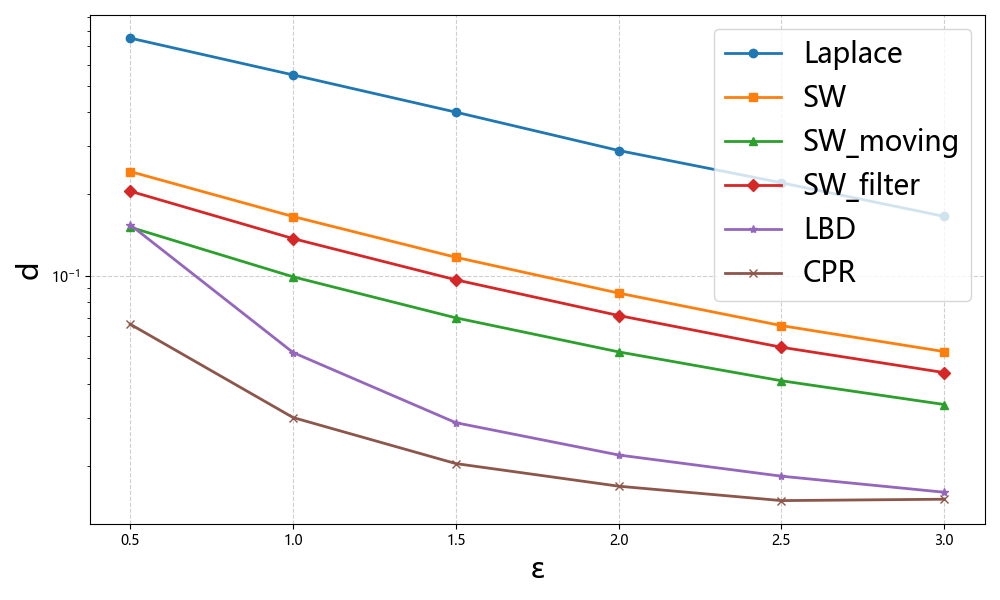}
        \caption{Raisin}
        \label{fig:raisin}
    \end{subfigure}
    \hfill
    % 第四张图: Crowdsourced
    \begin{subfigure}[b]{0.24\textwidth}
        \centering
        \includegraphics[width=\linewidth]{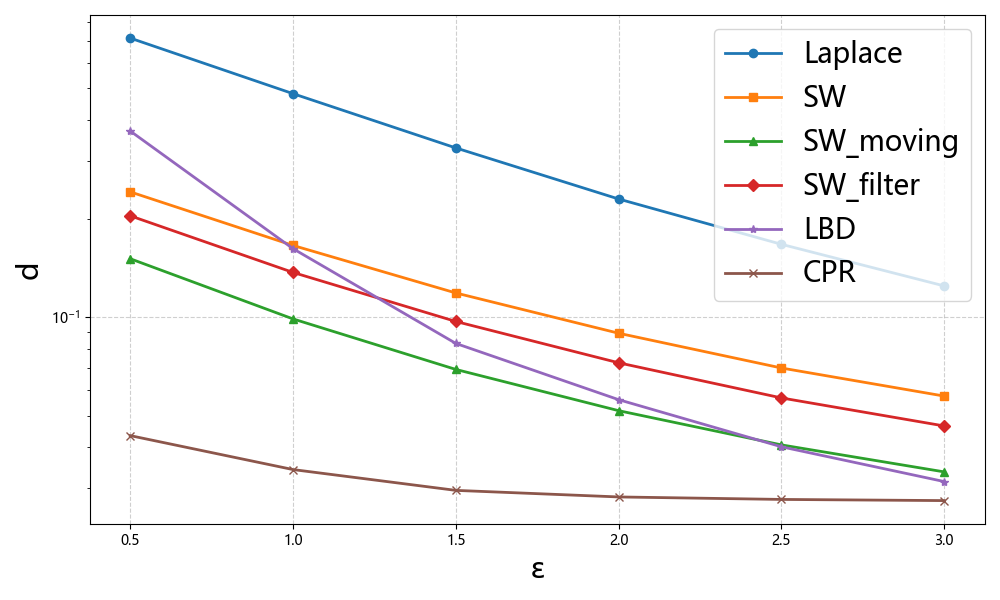}
        \caption{Crowdsourced}
        \label{fig:crowdsourced}
    \end{subfigure}
    
    \caption{Performance comparison of reconstruction methods across four datasets under various privacy budgets ($\epsilon$).}
    \label{fig:all_results}
\end{figure*}
As show in Figures \ref{fig:darwin},\ref{fig:music}, \ref{fig:raisin}, \ref{fig:crowdsourced}, CPR consistently achieves the highest reconstruction accuracy across all datasets and privacy budgets. While traditional Laplace and SW methods suffer from high LDP-induced variance, and LBD only matches our performance at large budgets, CPR maintains significant superiority, particularly under strict privacy constraints. Even smoothing-enhanced variants (SW+moving and SW filter) fail to bridge the gap, as they lack the periodicity-aware design essential for complex signals.

The primary strength of CPR stems from its \textbf{phase-aware statistical reconstruction}, as opposed to simplistic low-pass filtering. The low Cosine Distance confirms that our method preserves the structural morphology of the signal, thereby effectively avoiding the blurring of sharp transitions typical of moving-average approaches. By aggregating statistical phases across cycles, CPR reinforces feature consistency. Furthermore, our design specifically addresses two critical challenges in signal processing: \textbf{phase drift} and \textbf{boundary artifacts}. Mirror padding eliminates temporal discontinuities by providing symmetrical neighbor support, while phase-aligned aggregation circumvents temporal translation errors. Consequently, while maintaining the signal's intrinsic \textbf{temporal integrity}, CPR ensures that the reconstructed waveform retains essential peaks and timing characteristics, achieving robust privacy protection.

\section{Conclusion}
\label{4}
This paper addresses the challenges of period identification and time-series reconstruction for periodic data under LDP by proposing the \textbf{CPR} framework. CPR integrates multi-scale decomposition with multi-consensus detection to suppress spectral interference, and employs phase-sensitive robust aggregation to align observations across cycles, enabling high-precision reconstruction while respecting strict privacy budgets. Our experiments on real-world datasets demonstrate that CPR consistently preserves periodic structure and substantially improves reconstruction accuracy over existing methods.

\bibliographystyle{IEEEbib}
\bibliography{icme2026references}

\end{document}